\documentclass[usenatbib]{mn2e}
\usepackage{epsfig}
\usepackage{amsmath,amssymb}
\usepackage{natbib}
\usepackage{color}

\begin{document}

\title[The binary He-sdB  star: CPD$-20^{\circ}1123$]
{The helium-rich subdwarf CPD$-20^{\circ}1123$: 
a post-common envelope binary evolving onto the extended horizontal branch}

\author[Naslim, N., et al.]
{
Naslim~N.$^1$\thanks{E-mail: nas@arm.ac.uk}, 
S.~Geier$^2$\thanks{E-mail: Stephan.Geier@sternwarte.uni-erlangen.de},
C.~S.~Jeffery$^1$\thanks{E-mail: csj@arm.ac.uk},  
N.~T.~Behara$^{1,3}$,
V.~M.~Woolf$^4$ \&
L.~Classen$^2$\\
$^1$Armagh Observatory, College Hill, Armagh BT61\,9DG\\
$^2$Dr Karl Remeis-Sternwarte \& ECAP, Astronomisches Instit\"ut, 
Friedrich-Alexander Universit\"at Erlangen-N\"urnberg,\\ 
Sternwartstr. 7, D 96049 Bamberg, Germany\\
$^3$Institut d'Astronomie et d'Astrophysique, Universit\'e Libre de Bruxelles, Belgium\\
$^4$Physics Department, University of Nebraska at Omaha, Omaha, NE
68182-0266, USA
}

\date{Accepted .....
      Received ..... ;
      in original form .....}

\pagerange{\pageref{firstpage}--\pageref{lastpage}}
\pubyear{2012}
\label{firstpage}
\maketitle

\begin{abstract}
Subluminous B stars come in a variety of flavours including 
single stars, close and wide binaries, and pulsating and non-pulsating
variables. A majority have helium-poor surfaces (helium by
number $n_{\rm He}<1\%$), whilst a minority have extremely 
helium-rich surfaces ($n_{\rm He}>90\%$). 
A small number have an intermediate surface 
helium abundance ($\approx 10 - 30\%$), 
accompanied by peculiar abundances of other elements.
The questions posed are i) whether these abundance peculiarities 
are associated with radiatively-driven and time-dependent
stratification of elements within the photosphere as the star evolves 
from an helium-enriched progenitor to become a normal helium-poor
sdB star, and ii) whether these phenomena occur only in single sdB
stars or are also associated with sdB stars in binaries. 

We present a fine analysis of the bright 
intermediate-helium sdB star CPD$-20^{\circ}1123$ (Albus 1)
which shows it to be cool, for a hot subdwarf, 
with $T_{\rm eff}\approx23\,000$\,K and 
with a surface helium abundance $\approx17\%$ by number. 
Other elements do not show extraordinary anomalies; in common with 
majority sdB stars, carbon and oxygen are substantially depleted, 
whilst nitrogen is enriched. Magnesium through sulphur appear to be 
depleted by  $\approx0.5$\,dex, but chlorine and argon are 
substantially enhanced. 

We also present a series of radial-velocity measurements which show 
the star to be a close binary with an orbital period of 2.3\,d, 
suggesting it to be a post-common-envelope system. 

The discovery of an intermediate helium-rich sdB star in a close
binary in addition to known and apparently single exemplars 
supports the view that these are very young sdB stars in which 
radiatively-driven stratification of the photosphere is incomplete. 
\end{abstract}

\begin{keywords}
star: chemically peculiar (helium)
stars: evolution, 
stars: abundances,
stars: horizontal-branch,
stars: subdwarf,
stars: binary
\end{keywords}

\section{Introduction}
Subdwarf B stars are low-mass core helium burning stars with extremely
thin hydrogen envelopes. They behave as helium main-sequence stars of
roughly half a solar mass. Their atmospheres are generally helium
deficient; radiative levitation and gravitational settling combine to
make helium sink below the hydrogen-rich surface \citep{heber86}, to
deplete other light elements, and to enhance abundances of heavy
elements in the photosphere \citep{otoole06}.

However, almost 10$\%$ of the total subdwarf population comprise stars
with helium-rich atmospheres \citep{green86,ahmad06}. They have been
variously classified as sdOB, sdOC and sdOD \citep{green86} stars, but
more recently as He-sdB and He-sdO stars
\citep{moehler90,ahmad04}. Their optical spectra are characterised by
strong He\,{\sc i} (He-sdB) and He\,{\sc ii} (He-sdO) lines. The
helium-rich subdwarfs (He-sd's) may be further divided into {\it extremely} 
helium-rich stars ($\approx95\%$ of He-sd's 
having surface helium abundances $>80\%$ by number), and a small number of 
{\it intermediate} helium-rich stars ($\approx5\%$ of He-sd's, 
having surface helium abundances $>5\%$ and $<80\%$ by
number \citet{naslim10}). Whilst the extreme He-sdB stars may well be
the product of double helium white dwarf mergers \citep{zhang11}, the
intermediate He-sdB stars are more difficult to explain and include
stars as diverse as the prototype
JL\,87 \citep{ahmad07}, the extremely peculiar 
LS\,IV$-14^{\circ}116$ \citep{naslim11}, and also UVO\,0512--08 and
PG\,0909+276 \citep{edelmann03th}. 

CPD$-20^{\circ}1123$ (= Albus\,1 = TYC\,5940\,962\,1) is one of
the brightest known He-sdB stars with a V magnitude $11.75\pm0.07$
\citep{vennes07}. This object was first reported by \citet{gill96}
with a photographic magnitude 10.6. \citet{caballero07} suggested, it
might be a hot white dwarf or possibly a hot subdwarf at a distance
$d\approx 40$pc. \citet{vennes07} classified it as a bright
helium-rich subdwarf B star with $T_{\rm eff}=19\,800{\rm K}$ and
$\log g=4.55$ using low resolution optical spectra.

This paper reports a detailed atmospheric study 
 using a high-resolution optical spectrum
which confirms CPD$-20^{\circ}1123$ to be an {\it intermediate} 
helium-rich sdB star. It also 
reports a radial-velocity study which shows CPD$-20^{\circ}1123$ to be
the first  such star to be also a single-lined spectroscopic binary. 
It will be argued that these results may have profound implications for understanding 
the origin of {\it normal} sdB stars.

\section{Observations and Radial Velocities}

Service-mode observations of CPD$-20^{\circ}1123$ were made at the
Australian Astronomical Telescope (AAT) with the University College
London Echelle Spectrograph (UCLES) on 2010 January 14 using the 31
lines/mm grating to give a wavelength coverage of
$\lambda=3820-5230\,\mbox{\AA}$.  A series of 
six high-resolution optical spectra were made in very poor seeing
($>3"$) with 1 arc second slit delivering a resolution
$R\simeq45\,000$, and a total exposure time of 9000\,s.  The observations were
combined, flat-fielded, sky-subtracted and wavelength calibrated. 
The individual orders were merged and the final spectrum was normalized.

The AAT/UCLES spectrum displays strong C {\sc ii}, N {\sc ii}, Mg {\sc
  ii}, Al {\sc iii}, Si {\sc ii}, Si {\sc iii}, P {\sc iii}, S {\sc ii} and
S {\sc iii} lines, together with Ar {\sc ii}, Cl {\sc ii} and Fe {\sc iii}.
Unlike most other He-sdB stars, the optical spectrum shows relatively
strong hydrogen Balmer lines along with strong
Stark-broadened He {\sc i}\,4471, 4922, 4388 lines. This 
places the star in the list of hot subdwarfs with intermediate helium abundance,
alongside JL\,87 \citep{ahmad07} and LS\,IV$-14^{\circ}116$
\citep{naslim11}, but the absence of He{\sc ii} 4686\,\AA\ suggests a
much lower effective temperature. The lines due to chlorine, argon and
iron-group elements have not been observed in any other He-sdB star
\citep{naslim10}.

Fifteen medium-resolution spectra were obtained with the EMMI
spectrograph ($R\simeq3400,\lambda=3880-4380\,\mbox{\AA}$) mounted at
the European Southern Observatory's (ESO) New Technology Telescope
(NTT) in 2008 January.  Reduction was done with the ESO--MIDAS
package. The radial velocities (RV)
were measured by fitting a set of functions to the hydrogen Balmer
and neutral helium lines using the FITSB2 routine \citep{nap04}.
Gaussians were used to match the line cores, Lorentzians for the line wings and
polynomials to match the continuum. The RV of the star was found to be variable
on a timescale of days. The statistical $1\sigma$-errors of the single
measurements ranged from $5\,{\rm km\,s^{-1}}$ to $7\,{\rm km\,s^{-1}}$. In
order to derive the systematic uncertainties we calculated the standard
deviation of 8 RVs from spectra taken consecutively during the last night within
only $\simeq0.13\,{\rm d}$ assuming that the orbital period is much longer than
that. This standard deviation of $8\,{\rm km\,s^{-1}}$ was then adopted as the
uncertainty for all RVs.

Six additional high-resolution spectra have been taken with FEROS
($R\simeq48000,\lambda=3750-9200\,\mbox{\AA}$) mounted at the ESO
Max-Planck-Institut Garching (MPG) 2.2m telescope in 2010 October/November. 

The RVs were measured with high accuracy
from sharp, unblended metal and helium lines by fitting functions
with the Bamberg Spectrum Plotting and Analysis Suite SPAS \citep{Hirsch09}. 
The UCLES spectra were coadded and the RV determined in the same way. 
The RV measurements are given in Table~\ref{tab:RVs}.

\begin{table}
\caption{Radial velocities of CPD$-20^{\circ}1123$}
\label{tab:RVs}
\begin{center}
\begin{tabular}{lrl}
\hline
\noalign{\smallskip}
mid$-$HJD & RV [${\rm km\,s^{-1}}$] & Instrument\\
\noalign{\smallskip}
\hline
\noalign{\smallskip}
2454476.67364 & -31.8 $\pm$ 8.0 &  EMMI \\
2454476.72033 & -32.0 $\pm$ 8.0 &       \\
2454477.66701 &  16.3 $\pm$ 8.0 &       \\
2454477.83549 &  27.2 $\pm$ 8.0 &       \\
2454478.61086 &  15.9 $\pm$ 8.0 &       \\
2454478.66696 &  -4.3 $\pm$ 8.0 &       \\
2454478.75702 & -15.9 $\pm$ 8.0 &       \\
2454479.64573 & -39.8 $\pm$ 8.0 &       \\
2454479.65126 & -26.7 $\pm$ 8.0 &       \\
2454479.70277 & -21.2 $\pm$ 8.0 &       \\
2454479.70722 & -44.8 $\pm$ 8.0 &       \\
2454479.71159 & -31.9 $\pm$ 8.0 &       \\
2454479.76992 & -37.1 $\pm$ 8.0 &       \\
2454479.73092 & -26.5 $\pm$ 8.0 &       \\
2454479.78020 & -34.4 $\pm$ 8.0 &       \\
\noalign{\smallskip}                  
\hline                                
\noalign{\smallskip}                  
2455211.02431 & -11.5 $\pm$ 0.6 & UCLES \\
\noalign{\smallskip}                  
\hline                                
\noalign{\smallskip}                  
2455499.75901 & -12.1 $\pm$ 0.2 & FEROS \\
2455501.63676 &  31.1 $\pm$ 0.3 &       \\
2455501.64380 &  31.0 $\pm$ 0.3 &       \\
2455501.65652 &  29.7 $\pm$ 0.3 &       \\
2455501.82339 &  15.4 $\pm$ 0.2 &       \\
2455501.83216 &  14.9 $\pm$ 0.2 &       \\
\noalign{\smallskip}
\hline
\end{tabular}
\end{center}
\end{table}

\section{Orbital Solution}
\begin{figure}
	\centering
	\resizebox{\hsize}{!}{\includegraphics{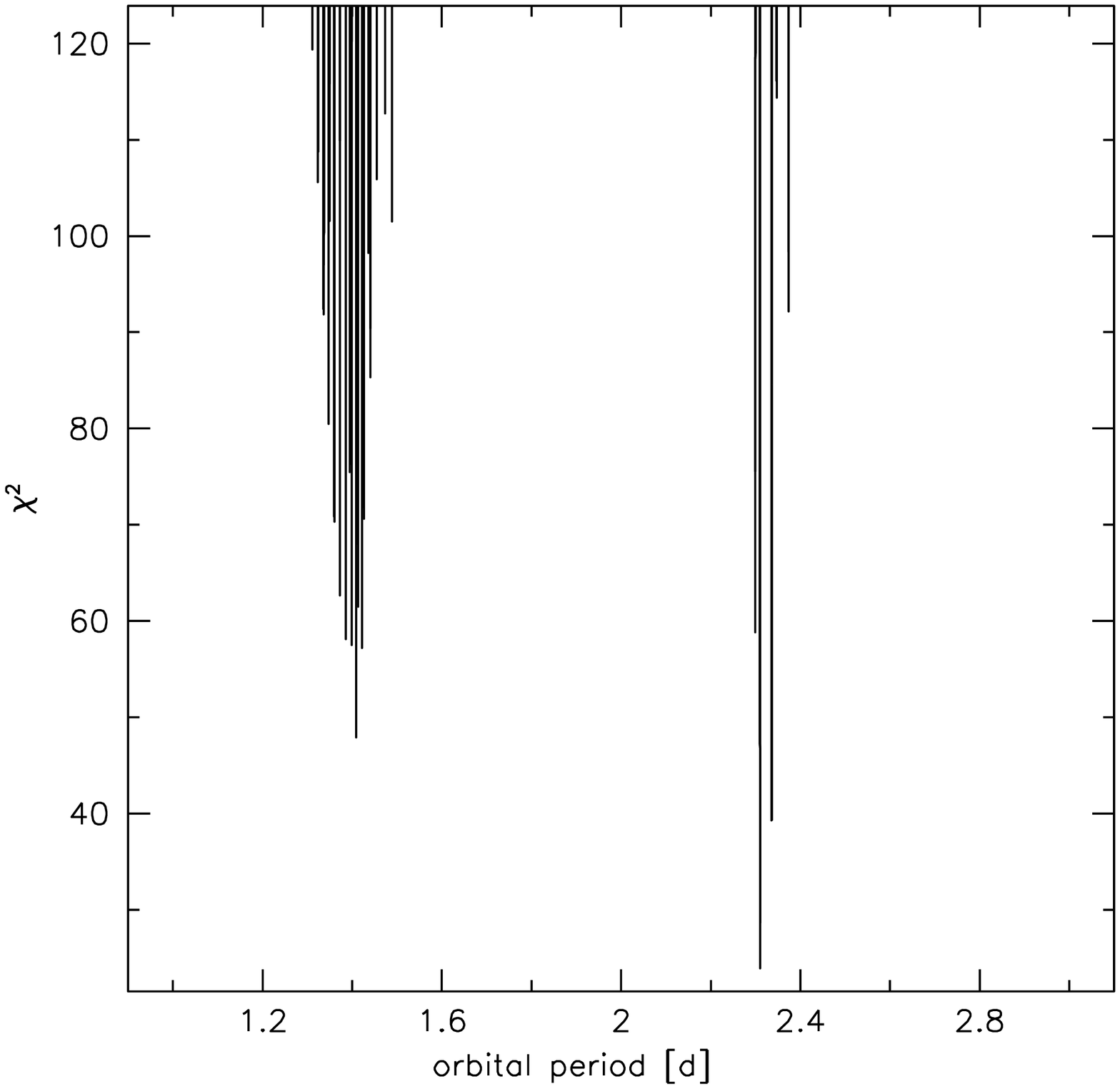}}
	\caption{$\chi^{2}$ of the best sine fit is plotted against
          the orbital period. Minima represent likely orbital
          periods. }
	\label{fig:chi}
\end{figure}

\begin{figure}
	\centering
	\resizebox{\hsize}{!}{\includegraphics{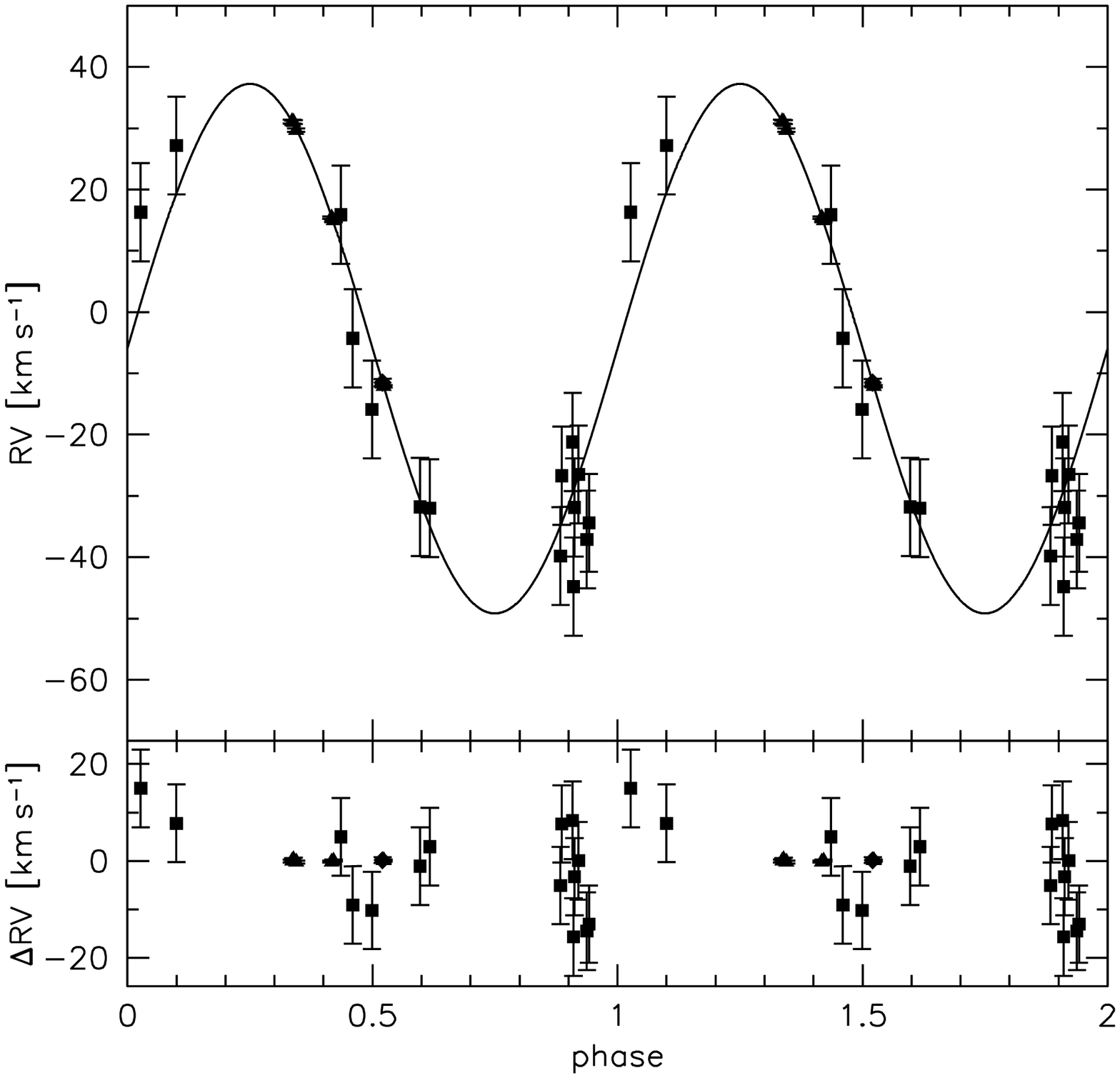}}
	\caption{Radial velocities of the subdwarf plotted against 
	orbital phase. The residuals are plotted below. Filled rectangles mark 
	RVs measured from EMMI spectra, filled triangles mark RVs obtained from 
	FEROS spectra and the filled diamond marks the RV measured
        from the coadded UCLES spectrum.}
	\label{fig:rv}
\end{figure} 
In order to derive the orbital solution, sine curves were fitted to the RV data
points in fine steps over a range of test periods. For each period the
$\chi^{2}$ of the best fitting sine curve was determined. The lowest $\chi^{2}$
indicates the most likely period. The orbital parameters are given in
Table~\ref{tab:par}.

We performed a Monte Carlo simulation for the most likely periods. For each of
the $10\,000$ iterations, a randomised set of RVs was drawn from Gaussian
distributions with central value and width corresponding to the RV measurements
and the analysis repeated (see Fig.~\ref{fig:chi}). The probability that the
solution with the lowest $\chi^{2}$ and $P=2.3098\pm0.0003\,{\rm d}$ is the
correct one is estimated to be $96\%$. The reduced $\chi^{2}$ of this solution
is $\simeq1.3$. The second best alias period ($P=2.3361\,{\rm d}$) with
$\Delta\chi^{2}=15$ has a probability of $0.8\%$ to be the correct one (see
Fig.~\ref{fig:chi}).

In order to derive conservative errors for the RV semi-amplitude $K$ and the
system velocity $\gamma$ we fixed the most likely period, created new RV
datasets with a bootstrapping algorithm and calculated the orbital solutions.
The standard deviation of these results was adopted as error estimate and is
about twice as high as the $1\sigma$-error. The phase folded RV curve is shown
in Fig.~\ref{fig:rv}, the orbital parameters are given in Table~\ref{tab:par}.

Adopting the frequently-assumed sdB mass of $0.47\,M_{\rm \odot}$, the minimum mass of
the companion ($0.21\,M_{\rm \odot}$) can be derived from the binary mass
function. No spectral features of the companion are visible. The unseen object
could be a late main-sequence star with a mass ranging from $0.21\,M_{\rm
\odot}$ to $\simeq0.45\,M_{\rm \odot}$. A more massive main-sequence star would
contribute sufficient light to be visible in the spectrum. The companion may
also be a white dwarf. In this case its mass would be less well constrained.

Owing to the long orbital period of CPD\,$-20^{\circ}1123$, it is hardly
possible to detect the reflection effect due to a cool M-dwarf companion with
ground-based photometry. \citet{koen09} detected the reflection effect with an
amplitude of 10\,mmag 
in the sdB+M-dwarf binary JL\,82. JL\,82 has an orbital period of 0.75\,d, making it
the 
longest-period sdB binary in which this effect has been observed. With an
orbital period three times longer, CPD\,$-20^{\circ}1123$ would exhibit a
reflection-effect  amplitude of a few mmag at most. Whilst such signals are 
detectable if the periods are sufficiently short (i.e. hours), 
they are almost impossible to detect if the periods are of the order of days.

\begin{table}
\caption{Orbital parameters of CPD$-20^{\circ}1123$} 
\label{tab:par}
\begin{center}
\begin{tabular}{ll}
        \noalign{\smallskip}
        \hline
        \noalign{\smallskip}
        $T_{\rm 0}$ [HJD]    & $2455500.86\pm0.01$ \\
        $P$                  & $2.3098\pm0.0003\,{\rm d}$ \\
        $\gamma$             & $-6.3\pm1.2\,{\rm km\,s^{-1}}$ \\
        $K$                  & $43.5\pm0.9\,{\rm km\,s^{-1}}$ \\
        $f(M)$               & $0.019\,M_{\rm \odot}$ \\
        $M_{\rm sdB}$ (adopted) & $0.47\,M_{\rm \odot}$ \\
        $M_{\rm 2,min}$      & $0.21\,M_{\rm \odot}$ \\
	\hline
\end{tabular}
\end{center}
\end{table}

\section{Atmospheric parameters}
\begin{figure}
\epsfig{file=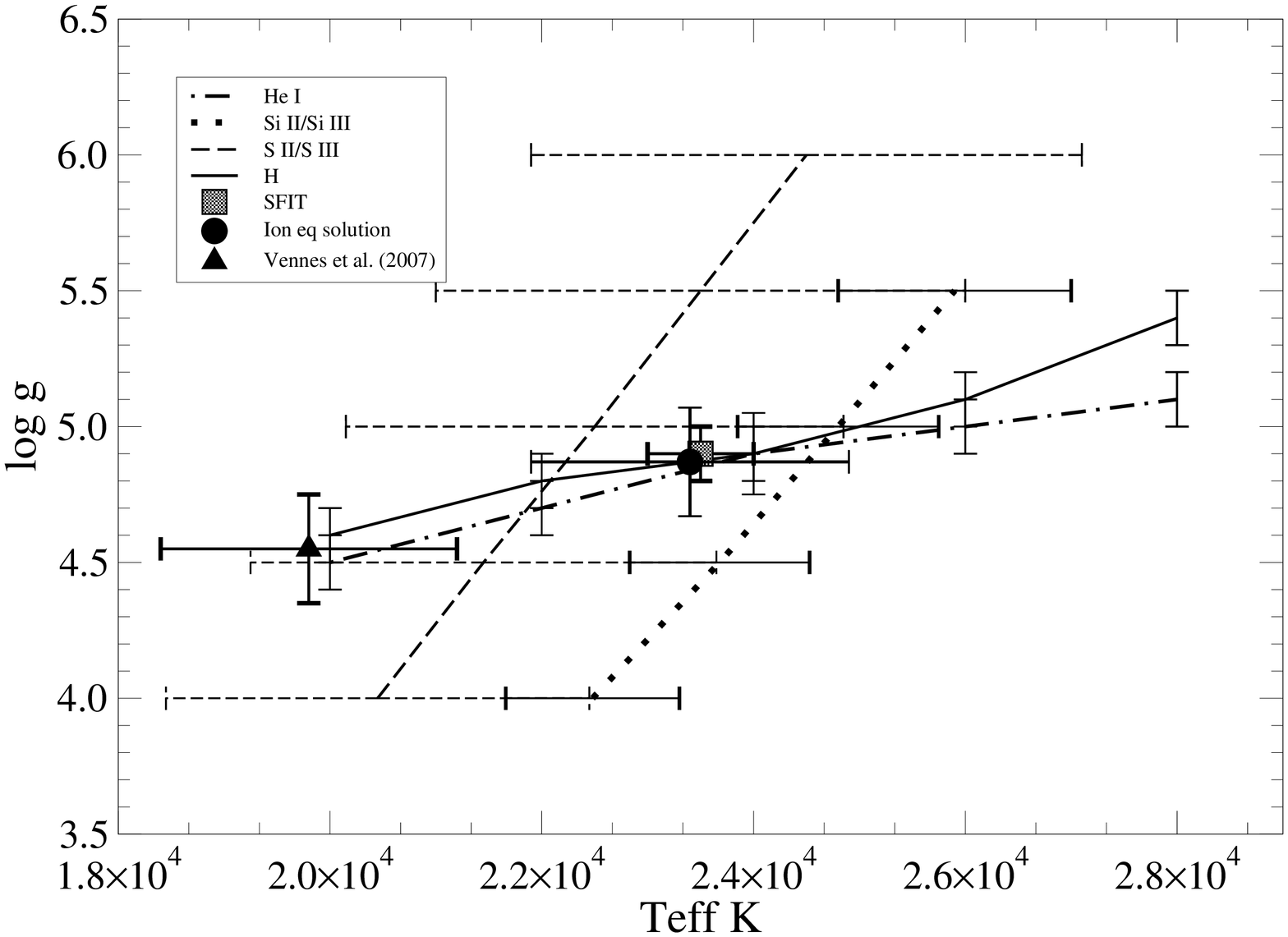, angle=-90, width=0.5\textwidth}
  \caption{The loci of ionisation equilibria for Si {\sc ii/iii}
    and S {\sc ii/iii}, the profile fits to H and He {\sc i} lines, and the
   adopted solution for CPD$-20^{\circ}1123$.} 
  \label{fig_ioneq}
\end{figure}
We measured the effective temperature $T_{\rm eff}$, surface gravity
$\log g$, and elemental abundances of CPD$-20^{\circ}1123$ using the
method described by \citet{naslim10}. The microturbulent velocity
determined by minimising the scatter in the nitrogen
abundance was $v_{\rm t}=9\pm3\,{\rm km\,s^{-1}}$. Since, no model
atmospheres with $v_{\rm t}=9\,{\rm km\,s^{-1}}$ were readily
available, those with $v_{\rm t}=10\,{\rm km\,s^{-1}}$ were used in
the subsequent formal solution for the ionisation equilibria and the
abundance measurements. We determined $T_{\rm eff}$ using the
ionisation equilibria of Si {\sc ii/iii} and S {\sc ii/iii}. The
surface gravity $\log g$ was determined by using line-profile fits to
the Stark-broadened He {\sc i} 4471 and Balmer (H$_{\beta}, \gamma$,
and $\delta$) lines. The coincidence of profile fits and ionisation
equilibria was used to determine the overall solution shown in
Fig.~\ref{fig_ioneq}. The ionisation equilibria and profile fits do
not coincide at a single point so that we took a weighted mean of the
intersections to determine $T_{\rm eff}$ and $\log g$.

In addition, the $\chi^{2}$-minimization package SFIT \citep{ahmad03}
was used to determine $T_{\rm eff}$, $\log g$ and $n_{\rm He}$
simultaneously. Initially we selected an assumed model atmosphere grid
with 1/10 solar metallicity as well as solar metallicity with $T_{\rm
  eff}=18\,000$ $(2000)$ $28\,000 {\rm K}$, $\log g = 4.00 (0.5) 5.50$
and $n_{\rm He}=0.2, 0.3, 0.5$. We adopted a microturbulent velocity
of $10\,{\rm km\,s^{-1}}$. Using SFIT we obtained a reduced $\chi^2$
fit for model with 1/10 solar metallicity and $n_{\rm He}=0.2$.  A
model atmosphere grid of $T_{\rm eff}=20\,000$ $(2000)$ $26\,000 {\rm
  K}$, $\log g = 4.00 (0.5) 5.50$ and $n_{\rm He}=0.1, 0.2, 0.3$ was
used to determine the best fit solution. The projected rotational
velocity obtained from the formal solution 
was $v \sin i \leqslant 1.0\pm0.5 {\rm km\,s^{-1}}$; the true value may
be slightly larger.

 The atmospheric parameters of CPD$-20^{\circ}1123$ measured using
 both SFIT and ionisation equilibrium agree with one another to within
 the formal errors and are shown in Table~\ref{t_pars}. Our estimates
 of $T_{\rm eff}$ and $\log g$ are different from the measurements by 
\citet{vennes07}. The latter used an intermediate resolution
 optical spectrum and line-blanketed NLTE model atmospheres for their
 analysis. They determined $T_{\rm eff}$, $\log g$ and $n_{\rm He}$
 from the Balmer and He {\sc i} lines fitting. The atmospheric
 parameters presented here were determined from Balmer and
 He {\sc i} lines fitting, Si {\sc ii/iii} and S
 {\sc ii/iii} ionisation equilibria. The reason for the discrepancy
 might be the use of two different methods of analyses. It is noted
 that, Si {\sc ii/iii} and S {\sc ii/iii} ionisation temperatures
 match better the Balmer and He {\sc i} lines at higher $T_{\rm eff}$
and $\log g$ (Fig.~\ref{fig_ioneq}).
\begin{table}
\centering
\caption{Atmospheric parameters of CPD$-20^{\circ}1123$}
\label{t_pars}
\begin{tabular}{@{}llll}
\hline
$T_{\rm eff} (\rm K)$& $\log g$      & $n_{\rm He}$ & Source\\
\hline
 $23\,500\pm500    $& $4.9\pm0.1$   & $0.170\pm0.05$ & SFIT \\
 $23\,400\pm1\,500 $& $4.87\pm0.2$ &       & Ion eq \\
 $19\,800\pm1\,400 $& $4.55\pm0.2$   & $0.150\pm0.15$   & \citet{vennes07} \\

\hline
\end{tabular}
\end{table}

\section{Abundances}

\begin{figure*}
\centering
\epsfig{file=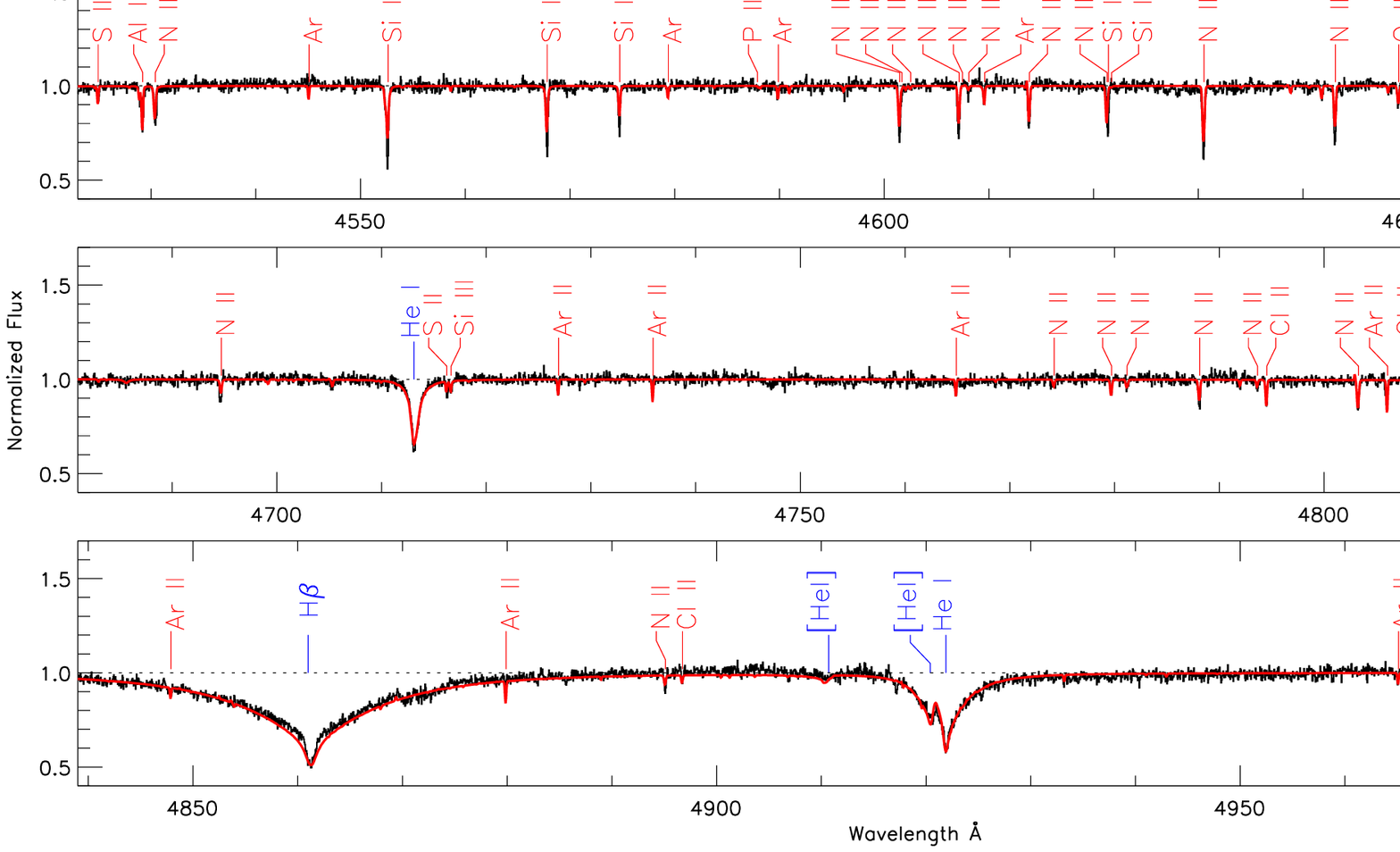,trim=1cm 3cm 1.5cm 0cm,width=15cm}
\caption{The merged AAT/UCLES spectrum of CPD$-20^{\circ}1123$ 
  along with the best FIT model
  of $T_{\rm eff}=24\,000\,{\rm K}$ and $\log g=5.0$. Abundances are as in Table~\ref{t_abunds1}.}
\label{fig_AATfit}
\end{figure*}

\begin{table*}
\caption{Individual line equivalent widths $W_{\lambda}$ and abundances $\epsilon_i$, with
 adopted oscillator strengths $gf$, for CPD$-20^{\circ}1123$.}
\begin{minipage}{8.5cm}
\centering
\label{t_lines1}
\begin{tabular}{@{}cccc}
\hline
Ion & & & \\
$\lambda/\mbox{\AA}$ &
$\log gf$ &
$W_{\lambda}/\mbox{m\AA}$ &
$\epsilon_{i}$  \\
\hline
C {\sc ii}      &           &       &           \\

4267.02 & $0.559\rceil$  &  58     & 6.71    \\
4267.27 & $0.734\rfloor$  &          &          \\

 N {\sc ii}       &           &       &         \\
3995.00 & 0.225  &  116  &   7.89      \\
4041.31 & 0.830  &  80  &   8.07     \\
4043.53 & 0.714  &  65   &  8.03       \\
4236.86 & $0.396\rceil$  &  85   &  8.19           \\
4236.98 & $0.567\rfloor$  &    &         \\
4241.78 & $0.728\rceil$  &  72  &  8.19     \\
4241.79 & $0.710\rfloor$  &    &           \\
4447.03 & 0.238  &  75   & 8.08       \\
4601.48 &$-0.385$&   85   & 8.34       \\
4607.16 &$-0.483$&   71   &  8.29      \\
4630.54 &  0.093 &  110   & 8.11        \\
4056.90 &$-0.461$&   32   &  8.76      \\
4073.05 & $-0.160$ &  42   & 8.61        \\

4171.59 & 0.281  &  32   & 8.06       \\
4176.16 &0.600&   53   & 8.05       \\
4530.40 &0.671&   51   &  8.12     \\
4643.09 &  $-0.385$ &  70   & 8.20        \\
4613.87 & $-0.607$ &  70   & 8.41        \\

 O {\sc ii}      &            &        &         \\
4649.14 & 0.342  &  26          &  7.53           \\
4414.90 & 0.210  &  10          &  7.17           \\
4069.88 &$0.365\rceil$&  12          &  7.26           \\

4069.62 & $0.158\rfloor$  &            &             \\
4075.86 & 0.700  &  21          &  7.47           \\
4416.97 &$-0.041$&  20          &  7.78           \\
Mg {\sc ii}      &      &        &                \\
4481.13 & $0.568\rceil$ &  137    &6.92    \\
4481.33 & $0.732\rfloor$ &     &    \\
  Al {\sc iii}      &            &        &         \\
4512.54&  0.405  &  45          &  6.26           \\
4529.20 &  0.660  &  55          &  6.12           \\
4479.97 &$1.021\rceil$ &  43          &  6.44           \\
4479.89 & $0.894\rfloor$  &            &             \\
4149.90 & $0.619\rceil$  &  43          &  6.29         \\
4150.14 &$0.464\rfloor$&          &           \\
 Si {\sc ii}                    &          &           \\
 4128.07&  0.369  &  50          &  6.90           \\
\hline
\end{tabular}\\

\end{minipage}
\begin{minipage}{8.5cm}
\label{t_lines}
\begin{tabular}{@{}cccc}
\hline
Ion & & & \\
$\lambda/\mbox{\AA}$ &
$\log gf$ &
$W_{\lambda}/\mbox{m\AA}$ &
$\epsilon_{i}$  \\
\hline
     &            &        &         \\

4130.89 &  0.545  &  51          &  6.73           \\

Si {\sc iii}      &            &        &         \\

4552.62 & 0.283 &  111   & 6.96   \\
4567.82 & 0.061 &  85   & 6.96   \\
4574.76 & $-0.416$ &  60   & 7.17   \\
 P {\sc iii}      &            &        &         \\
4059.31&  $-0.050$  &  22          &  5.62          \\
4222.19 &   0.190  &  40          &  5.74           \\

S {\sc ii}      &            &        &         \\
4815.52 &  $-0.050$  &  47          &  7.44           \\
4716.23 &$-0.050$ &  22          &  7.50           \\
4524.95 &$0.061$  &  25            &  7.36          \\
4294.43 &0.560    &  30            &  7.20           \\
4162.70 & 0.785  &33            &  6.97          \\
4153.10 & 0.681  & 33          &  7.08          \\

S {\sc iii}      &            &        &         \\
4253.59 & 0.233 &  36   & 6.72   \\
4284.99 & $-0.046$ &  30   & 7.03   \\
Cl {\sc ii}      &            &        &         \\
4794.54 & 0.423 &  31   & 6.38   \\
4810.06 & 0.281 &  25   & 6.43   \\
4896.74 & 0.450 &  28   & 6.96$\dagger$ \\

 Ar {\sc ii}       &           &       &         \\
4401.02 & $-0.250$  &  16  &   6.91      \\
4371.36 & $-0.570$  &  25  &   7.44     \\
4806.07 & 0.215  &  37   &  6.97      \\
4879.90 & $0.220$  &  24   & 6.88            \\
4426.01 & $0.170$  & 26   &  6.82       \\
4609.60 & $0.286$  &  21  &  7.05     \\
4726.91 & $-0.180$  & 15   & 7.03          \\
4735.93 & $-0.108$  &  24   & 7.06      \\
4764.89 &$-0.110$&   15   & 7.00       \\
4657.90 &$-0.283$&   14   &  7.10     \\
Fe {\sc iii}      &            &        &         \\
4164.73 & 0.935 &  24   & 7.13   \\
4166.84 & 0.436 &  15   & 7.38   \\
4419.60& $-2.218$ &  20   & 7.16  \\
4395.76& $-2.595$       &  12     &7.30  \\
4166.86& 0.436       &  15     &7.38  \\
                     &         &      \\
\hline
\end{tabular}
\end{minipage}\\
\parbox{170mm}{
$gf$ values:
C {\sc ii} \citet{yan87}, 
N {\sc ii} \citet{bec89}, 
O {\sc ii} \citet{Bec88},
Mg {\sc ii} \citet{wie66}
Al {\sc iii} \citet{top92,McE83}, 
Si {\sc ii} \citet{bec90},
Si {\sc iii} \citet{bec90},
P {\sc iii} \citet{wie69},
S {\sc ii} \citet{wie69},
S {\sc iii} \citet{wie69, har70},
Cl {\sc ii} \citet{rod89},
Ar {\sc ii} \citet{wie69},
Fe {\sc iii} \citet{kur91}.
} \\
\parbox{170mm}{$\dagger$: line abundance omitted from mean in Table 
\ref{t_abunds1} }
\end{table*}

\begin{table*}
\centering
\caption{Mean abundances $\epsilon_i$ for CPD$-20^{\circ}1123$.}
\label{t_abunds1}
\begin{tabular}{@{}ccccccc}
\hline
El. &CPD$-20^{\circ}1123$  & sdB$^{1-5}$ & JL\,87$^6$ & LS\,IV$-14^{\circ}116^7$    &He-sdB$^8$ & Sun$^9$ \\
\hline
H   & $11.85\pm0.1$&   12.0   & 11.6 & 11.83     &$<8.5-11.1$ & 12.00  \\
 He  & $11.17\pm0.1$& 7.9--11.0& 11.3 & 11.15     &11.5 & [10.93]  \\
 C   & $6.71\pm0.1$& 5.5--9.5 &  8.8 &  8.04     &6.5--9.0  &  8.52  \\
 N   & $8.21\pm0.21$& 6.5--8.5 &  8.8 & 8.02     &8.0--9.0  &  7.92  \\
 O   & $7.41\pm0.23$& 6.0--8.5 &  8.6 & 7.60     &6.8--7.5  &  8.83  \\
 Ne  & $<7.5 $      & 6.5--8.5 & 8.31   & $<7.6$   &7.7--9.0 &  [8.08]  \\
 Mg  & $6.92\pm0.10$& 5.5--7.8 &  7.4 &  6.85    &7.0--8.5  &  7.58  \\
 Al  & $6.31\pm0.13$&4.5--7.0  &6.3   & &6.0--6.4  & 6.47   \\
 Si  & $6.94\pm0.16$& 5.0--7.7 &  7.2 &  6.32    &6.5--7.5  &  7.55  \\
 P   & $5.68\pm0.10$&4.5--6.0  & 5.3  & &  & 5.45    \\
 S   & $7.09\pm0.28$&5.0--8.0        &  6.9    &      &  6.0--7.0   &  7.33      \\
Cl   & $6.40\pm0.10$&                &         &      &             &  5.5        \\
Ar   & $7.03\pm0.17 $          & 6.0--9.0       &  6.3    & $<6.5$   &       &  [6.40]      \\
Sc   & $<3.5 $          & 5.0--7.0 &      &  $<5.3$    &     &  3.17      \\
Ti   & $<5.8 $          & 5.3--9.0 &      &   $<6.0$   &     &  5.02      \\
V    & $<6.0  $         & 6.0--8.5 &      &  $<6.5$    &     &  4.00      \\
Cr   & $<6.5  $         & 5.5--8.0 &      &  $<7.0$    &     &  5.67      \\
Fe   & $7.27\pm0.12 $          & 6.5--8.1 & 7.5     & $<6.8$    &      &  7.50      \\
\hline
\end{tabular}\\
\parbox{150mm}{
References: 1. \citet{edelmann03},
2. \citet{geier08},
3. \citet{geier10},
4. \citet{otoole06}, 
5. \citet{chayer06},
6. \citet{ahmad07},
7. \citet{naslim11},
8. \citet{naslim10},
9. \citet{grevesse98}.}
\parbox{150mm}{
The solar helium abundance is the
asteroseismic value for the outer convection zone, the solar neon and argon
abundances are the meteoritic value; other solar abundances are for the solar
photosphere.
}
\end{table*}

For abundance measurements the model atmosphere with $T_{\rm
  eff}=24\,000\,{\rm K}$, $\log g=5.0$, $n_{\rm He}=0.2$ and 1/10
solar metallicity was adopted.  After measuring the equivalent widths
of all C, N, O, Mg, Al, Si, S, Cl, Ar and Fe lines using the spectrum
analysis tool DIPSO, the abundances were calculated using 
the LTE radiative transfer code SPECTRUM \citep{jeffery01}.
The adopted oscillator strengths ($gf$), equivalent widths and lines
abundances are given in Table~\ref{t_lines1}. Abundances are given in
the form $\epsilon_i = \log n_i + c$ where $\log \Sigma_i a_i n_i =
12.15$ and $a_i$ are atomic weights. This form conserves values of
$\epsilon_i$ for elements whose abundances do not change, even when
the mean atomic mass of the mixture changes substantially.

Mean abundances for each element are given in
Table~\ref{t_abunds1}. The errors given in Table~\ref{t_abunds1} are
based on the standard deviation of the line abundances about the mean
or in the case of a single representative line, the estimated error in
the equivalent width measurement. Systematic shifts attributable to
errors in $T_{\rm eff}$ and $\log g$ are given in
Table~\ref{t_err1}. The line abundance for Cl\,{\sc ii}\,4896.7\AA\ is 
conservatively omitted from the mean since the high abundance implied
is not supported by the non-detection of Cl\,{\sc ii}\,4904.7\AA. 

The final best-fit spectrum using the adopted
best-fit model and the elemental abundances from
Table~\ref{t_abunds1} is shown in Fig.~\ref{fig_AATfit}, together with
identifications for all of the absorption lines in the model. The
elemental abundances shown in Table~\ref{t_abunds1} are the mean
abundances of all individual lines of an ion. Consequently, 
the cores of strong and saturated lines do not fit perfectly
in Fig.~\ref{fig_AATfit}.
A few features on the observed spectrum (e.g. 4423.93\,\AA) 
are not explained by the model, but might be defects from
the order merging process. 

Table~\ref{t_abunds1} also compares the representative range of
abundances measured for normal sdB stars, for intermediate helium sdB
stars JL\,87 and LS\,IV$-14^{\circ}116$, for extreme helium-rich sdB
stars and the Sun. CPD$-20^{\circ}1123$ appears to be metal poor
($\approx 0.5\pm0.3$) using magnesium, aluminium, silicon, sulphur and
iron as proxies for overall metallicity. The nitrogen abundance is
nearly solar and carbon is underabundant by $\approx 1.8$ dex. This
star shows strong argon and chlorine enrichment ($\approx 0.63$ and
0.9 dex, respectively). 
The iron group elements Sc,
Ti, V, Cr and Fe have been reported in certain normal sdB stars
\citep{geier08,geier10,otoole06}. The iron abundance in
CPD$-20^{\circ}1123$ is found to be subsolar and no detectable lines
due to Sc, Ti, V and Cr were identified in the optical spectrum. The
upper limit abundances of these elements were determined, assuming 
equivalent widths $<5{\rm m\AA}$ for the strongest lines of these 
species (Table~\ref{t_abunds1}).  More lines due to Fe and other iron
group elements should be observable in the ultraviolet. The UCLES
data used for this analysis are insufficient to search for variability.

\begin{table}
\centering
\caption{Systematic abundance errors $\delta \epsilon_i$ due to representative errors
  in $T_{\rm eff}$ and $\log g$.}
\label{t_err1}
\begin{tabular}{@{}lll}
\hline
Element& $\delta T_{\rm eff}=1000\,\rm K$     & $\delta \log g=0.2$  \\
\hline
 C   & 0.05   &   --0.03      \\
 N   & 0.15     &  --0.06    \\
 O   & 0.18     &  --0.08       \\
 Mg  & --0.09      & 0.03      \\
 Al  & 0.11      & --0.05      \\
Si  &  0.05    &  --0.04       \\
P    & 0.11          &  --0.08           \\
 S  &  0.05    &  --0.04       \\
Cl  &  --0.20   &   0.05       \\    
 Ar  & 0.005      & --0.02        \\
 Fe  & 0.11    & --0.08        \\

\hline
\end{tabular}
\end{table}
\begin{figure}
\epsfig{file=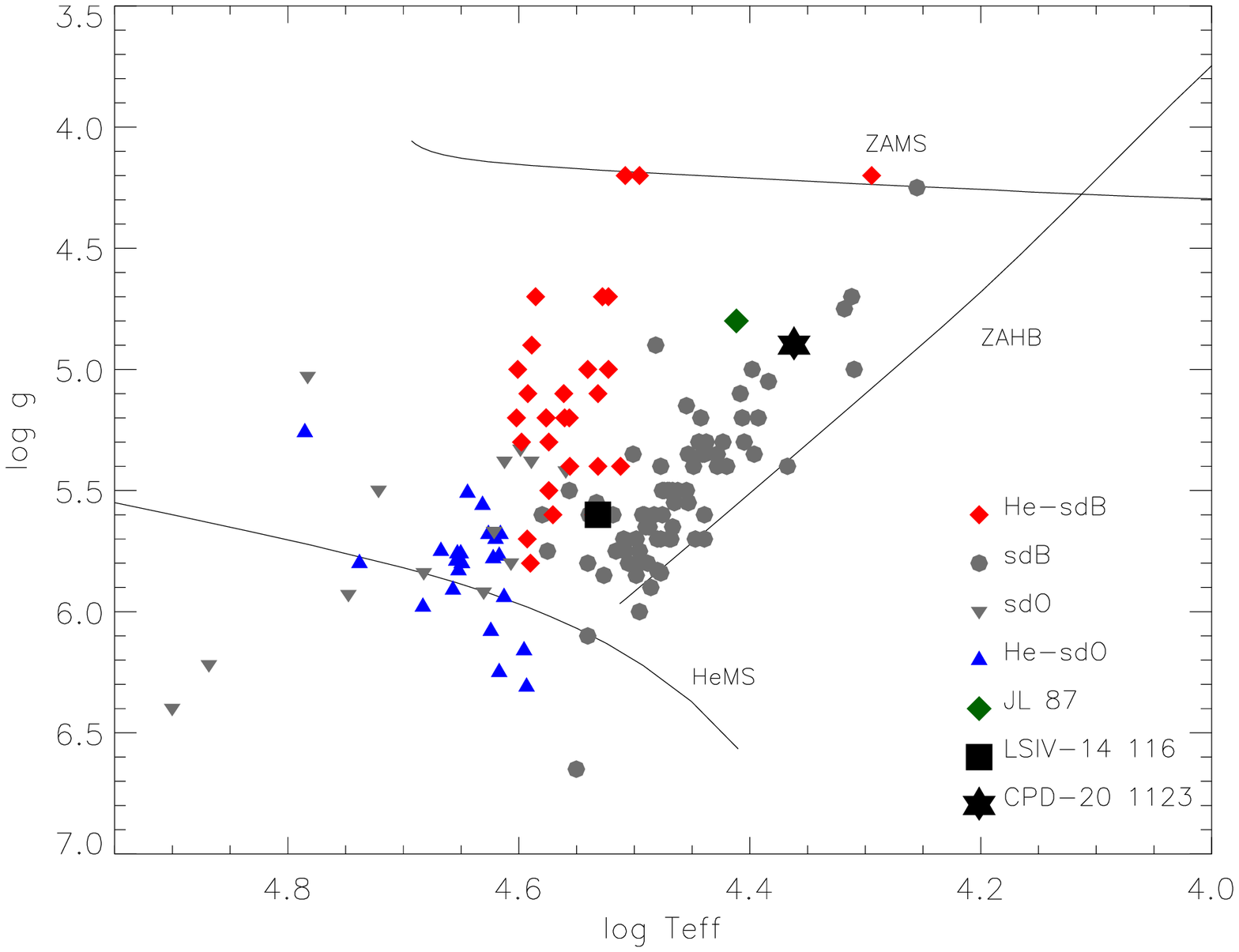, angle=0, width=0.5\textwidth}
  \caption{The location of CPD$-20^{\circ}1123$ on the $\log g -
    T_{\rm eff}$ diagram, compared with normal sdB stars \citep{edelmann03}
    helium-rich sdB stars \citep{ahmad03,naslim10}, helium-rich sdO
    stars \citep{stroeer07}, LS\,IV$-14^{\circ}116$ \citep{naslim11} and JL\,87 \citep{ahmad07}. A zero-age
    main sequence ($Y=0.28,Z=0.02$), a zero-age horizontal branch
    ($M_{\rm c}=0.485 {\rm M_{\odot}}, Y_{\rm e}=0.28, Z=0.02$) and a
    helium main sequence ($Z=0.02$) are also shown.} 
  \label{teff-log1}  
\end{figure}
\section{Is CPD\,$-20^{\circ}1123$ a Helium-Rich Star?}

CPD\,$-20^{\circ}1123$ attracted attention because it appears to be a
bright sdB star with unusually strong helium lines
\citep{vennes07}. Indeed, a fine analysis demonstrates the helium
abundance to be nearly a factor of two greater than solar. This places CPD\,$-20^{\circ}1123$ firmly amongst the
intermediate helium-rich sdB stars alongside JL\,87,
LS\,IV$-14^{\circ}116$, and others. A {\it normal} sdB star tends to have helium abundances 2 or 3 dex below solar.

The question raised by the intermediate helium-rich sdB stars is how
or whether they are related to normal sdB stars, or the extreme
helium-rich sdB stars, or to neither. One is tempted to distinguish
the extreme helium-rich and normal sdB stars on the basis of position
in the $g-T_{\rm eff}$ diagram (Fig.~\ref{teff-log1}) as well as
surface composition. \citet{zhang11} argue strongly that the former
are post double-white dwarf mergers evolving onto the helium
main-sequence, with surface chemistries dominated by the constituents
of the merger.

The intermediate helium-rich sdB stars, on the other hand, lie much
closer to the locus of the normal sdB stars although also toward the
low-gravity boundary.  It is important to remember that the surface
chemistry of normal sdB stars is dominated by diffusion activated by
strong gradients in the radiative forces acting on different
elements. The chemical separation induced by these gradients does not
occur instantaneously but takes some 10$^5$ to 10$^6$ years, a small
but finite fraction of the lifetime of an extended horizontal-branch
star ($10^8$ y).

What is the surface chemistry of an sdB star progenitor?  Clearly, the
answer depends on what sort of star the progenitor was. Excluding
double-white dwarf mergers, current understanding indicates the 
most likely possibilities are helium-core
red giants \citep{dorman93} which lose their outer layers
either by a fast stellar wind \citep{Dcruz96}, by stable Roche lobe
overflow \citep{green00}, or by common-envelope ejection
\citep{han02}. As a consequence of first dredge-up on the first giant
branch, it is expected that the surface layers of the red giant will
have above-solar abundances of helium and nitrogen.  
Since less than $0.002
{\rm M_{\odot}}$ of hydrogen-rich material survives the mass-loss
episode, which is less than the natural thickness of the
hydrogen-burning shell in a red giant, 
these remaining layers may or may not be deep enough to have been further
enriched in helium. Following core-helium ignition, the surviving star
contracts onto the extended-horizontal branch by one of several routes
 \citep{lanz04,miller08,miller11} on a timescale of $\approx$ 10$^6$ y.
 
Once the surface temperature exceeds some 10\,000\,K and the the
surface gravity approaches some 10\,000\,cm\,s$^{-2}$,
radiatively-driven diffusion becomes effective, and ultimately the
helium and other light elements sink beneath the hydrogen. During this
process, it is not inconceivable that the outer layers of the star
become heavily stratified, with elements being concentrated at depths
where their specific opacities are high. Such appears to be the case
for the extraordinary "zirconium star" LS\,IV$-14^{\circ}116$. Such
exotic mixtures may simply be a precursor to that which is referred to
as normal for an sdB star.

Now it is well-established that some 50 -- 70 \% (at least) of normal
sdB stars are members of binary systems \citep{maxted01,geier11}.  If
intermediate helium-rich sdB stars are simply the precursors of normal
sdB stars, then the fraction of both groups which are binaries should
be the same.  Until now, an argument against the precursor hypothesis
was that no intermediate helium star was known to be a binary.
However, such an argument only has merit with the inclusion of 
UVO\,0512--08 and PG\,0909+276 \citep{edelmann03th} in addition 
to LS\,IV$-14^{\circ}116$ \citep{naslim11} and JL\,87 \citep{ahmad07}.  
The discovery that
CPD$-20^{\circ}1123$ is a short-period binary with a probable
white-dwarf or a late main-sequence companion reverses the argument, and supports the possibility that
intermediate helium-rich and normal sdB stars derive from a similar
group of progenitors.  We note this is {\it not} the same as saying
that one type evolves into the other.

Two additional facts may be important. \\ 
i) All known intermediate helium-rich sdB stars\footnote{JL\,87,
  LS\,IV$-14^{\circ}116$, UVO\,0512--08, PG\,0909+276 and
  CPD$-20^{\circ}1123$} 
have very small projected rotation velocities ($v \sin i$).  A
similar result has been found for normal sdB stars which are {\it not}
in close binaries with periods less than $\approx 1.2$ days
\citep{geier12}. Slow rotation is believed to be a necessary condition
contributing to chemical peculiarity in mercury-manganese stars \citep{wolff78}.
Fast rotation may cause 
mixing effects which prevent elements from being concentrated in 
stratified layers.
\\ 
ii) CPD$-20^{\circ}1123$ lies at the cool end of the sdB domain. The
diffusion time-scale should be longer at lower $T_{\rm eff}$ and $g$,
so that peculiar surface chemistries may be relatively more likely. It
would be interesting to look for more helium-strong sdB stars at the
cool end of the sdB sequence.

\section{Conclusion}

With a helium abundance about twice solar, CPD\,$-20^{\circ}1123$ is
securely determined to be an intermediate helium-rich subdwarf B
star. Chemically these are quite distinct from normal sdB stars (which
are helium poor) and extremely helium-rich sdB stars (more than 80\%
helium). The abundances of other elements in CPD\,$-20^{\circ}1123$
are slightly peculiar; high argon and chlorine, and detectable 
iron are noted in particular.
 
 CPD\,$-20^{\circ}1123$ is the first intermediate helium-rich sdB star
 found to be a short-period binary. With a period of 2.3\,d, the
 companion could be either a low-mass main-sequence star or a white
 dwarf.

It is argued that a high helium abundance would be expected for the
surface of a very young sdB star, evolving from the tip of the first
red-giant branch after having lost its hydrogen-rich envelope. This
helium will sink out of sight within the first $10^6$ years of
evolution on the extended horizontal branch. It is proposed that the
intermediate helium sdB stars are simply very young normal sdB stars
in which this process has yet to be completed.

It is expected that as the number of known intermediate-helium sdB
stars increases, the number found to be members of binaries will also
increase, as will the variety of their surface chemistries.
Understanding how the surface chemistries of these stars change is a
challenge for both theory and observation but will ultimately help to
explain how sdB stars are formed.

\section*{Acknowledgments}

This paper is based on observations obtained with the Australian Astronomical 
Telescope and on observations at the La Silla Observatory of the European Southern Observatory for programmes number 080.D-0685(A) and 086.D-0714(A).

The Armagh Observatory is funded by direct grant from the Northern
Ireland Dept of Culture Arts and Leisure. 
S.~G. is supported by the Deutsche Forschungs\-gemeinschaft under grant HE1356/49-1.

\bibliographystyle{mn2e}
\bibliography{mnemonic,CPDS201123}

\label{lastpage}




\end{document}